\documentclass{article}
\usepackage{spconf,amsmath,graphicx}
\usepackage[pagebackref=true]{hyperref}
\usepackage{amsmath,amsfonts,amsthm,amssymb,booktabs}
\usepackage{enumitem}
\usepackage{caption,subcaption}
\usepackage{multicol,multirow}
\usepackage{algorithm,algpseudocode}
\usepackage{xcolor}
\usepackage{pifont}
\usepackage{graphicx}
\usepackage{bm}
\usepackage[normalem]{ulem}

\newcommand{\cmark}{{\color{blue}\ding{51}}}%
\newcommand{\xmark}{{\color{red}\ding{55}}}%

\newcommand{\rotthirty}[1]{\rotatebox[origin=c]{30}{#1}}

\newcommand{\rotninety}[1]{\rotatebox[origin=c]{90}{#1}}

\usepackage{arydshln}

\newcommand{\var}[3]{\mathrm{#1}_{#2}^{#3}}

\renewcommand{\vec}[3]{\mathbf{#1}_{\mathrm{#2}}^{\mathrm{#3}}}
\newcommand{\bvec}[3]{\boldsymbol{#1}_{\mathrm{#2}}^{\mathrm{#3}}}

\newcommand{\subsup}[3]{#1_{#2}^{#3}}
\newcommand{\subsub}[2]{#1_{_{#2}}}

\newcommand{\abs}[1]{\left|#1\right|}

\newcommand{\eqn}[1]{{\begin{equation}#1\end{equation}}}

\newcommand{\cleanhundred}{{$\mathrm{clean}\text{-}100$}}
\newcommand{\cleanthreesixty}{{$\mathrm{clean}\text{-}360$}}
\newcommand{\cleanfoursixty}{{$\mathrm{clean}\text{-}460$}}
\newcommand{\otherfivehundred}{{$\mathrm{other}\text{-}500$}}
\newcommand{\mixedeightsixty}{{$\mathrm{mixed}\text{-}860$}}
\newcommand{\allninesixty}{{$\mathrm{all}\text{-}960$}}

\newcommand{\devclean}{$\mathrm{dev}\text{-}\mathrm{clean}$}
\newcommand{\devother}{$\mathrm{dev}\text{-}\mathrm{other}$}
\newcommand{\testclean}{$\mathrm{test}\text{-}\mathrm{clean}$}
\newcommand{\testother}{$\mathrm{test}\text{-}\mathrm{other}$}

\newcommand{\cohortsize}{$\mathrm{cohort\_size}$}
\newcommand{\localepochs}{$\mathrm{local\_epochs}$}
\newcommand{\aggrounds}{$\mathrm{aggregation\_rounds}$}
\newcommand{\optswitch}{$\mathrm{OptSwitch}$}

\newcommand{\flrandomstart}{$\mathrm{FL}\text{-}\mathrm{random}\text{-}\mathrm{start}$}
\newcommand{\flseedstart}{$\mathrm{FL}\text{-}\mathrm{seed}\text{-}\mathrm{start}$}
\newcommand{\attentionrescore}{$\mathrm{attention\_rescoring}$}
\newcommand{\stddev}{$\mathrm{std}$}
\newcommand{\flforasr}{$\mathrm{FL4ASR}$}
\newcommand{\adam}{$\mathrm{Adam}$}
\newcommand{\sgd}{$\mathrm{SGD}$}
\newcommand{\adamw}{$\mathrm{AdamW}$}
\newcommand{\yogi}{$\mathrm{Yogi}$}
\newcommand{\lamb}{$\mathrm{LAMB}$}
\newcommand{\warmstart}{$\mathrm{ws}$}
\newcommand{\pretrained}{$\mathrm{pt}$}
\newcommand{\warmup}{$\mathrm{wu}$}
\newcommand{\decay}{$\mathrm{dec}$}
\newcommand{\wer}{\mathrm{WER}}

\newcommand{\diverges}{$\mathrm{training\,does\,not\,converge}$}

\hypersetup{
    colorlinks=true,
    linkcolor=blue,
    citecolor=blue,
    urlcolor=blue
}

\title{Importance of Smoothness Induced by Optimizers in FL4ASR: \\Towards Understanding Federated Learning for End-to-End ASR}
\name{Sheikh Shams Azam, Tatiana Likhomanenko, Martin Pelikan, Jan ``Honza'' Silovsky}
\address{Apple}
\begin{document}
\maketitle
\begin{abstract}
In this paper, we start by training End-to-End Automatic Speech Recognition (ASR) models using Federated Learning (FL) and examining the fundamental considerations that can be pivotal in minimizing the performance gap in terms of word error rate between models trained using FL versus their centralized counterpart. Specifically, we study the effect of (i) adaptive optimizers, (ii) loss characteristics via altering Connectionist Temporal Classification (CTC) weight, (iii) model initialization through seed start, (iv) carrying over modeling setup from experiences in centralized training to FL, e.g., pre-layer or post-layer normalization, and (v) FL-specific hyperparameters, such as number of local epochs, client sampling size, and learning rate scheduler, specifically for ASR under heterogeneous data distribution. We shed light on how some optimizers work better than others via inducing smoothness. We also summarize the applicability of algorithms, trends, and propose best practices from prior works in FL (in general) toward End-to-End ASR models.  
\end{abstract}
%

\section{Introduction}
\label{sec:intro}

In recent years, End-to-End (E2E) Automatic Speech Recognition (ASR) model training using Connectionist Temporal Classification (CTC) \cite{graves2006connectionist}, Transducer~\cite{graves2012sequence}, or Attention-based Encoder-Decoder (AED)~\cite{chorowski2015attention} losses has become ubiquitous because of the simplified training/inference procedures and state-of-the-art (SOTA) performance. For example, WeNet \cite{yao2021wenet} is a recently proposed E2E ASR training library aimed towards production-first development owing to the growing interest in the deployment of these models for practical use cases. These E2E ASR models -- constituting Transformer~\cite{vaswani2017attention} or Conformer~\cite{gulati2020conformer} blocks in various configurations -- are often several orders of magnitude larger than the models that were trained using a conventional hybrid ASR framework, are trained on large servers for several days, and require much larger datasets that often are acquired through data aggregation from opt-in users to reach SOTA performance.

\vspace{1mm}
In parallel to these strides in ASR, there has been an increasing push towards developing algorithms and techniques that require less data, prevent privacy leakage~\cite{azam2022can}, or provide users with privacy guarantees~\cite{dwork2014algorithmic} over the aggregated datasets. Federated Learning (FL) is one such technique that decentralizes model training so that data do not have to be stored on the server. It can also incorporate privacy guarantees, e.g., differential privacy \cite{brendan2018learning} or secure aggregation~\cite{bonawitz2017practical}. FL thus allows the training of models while protecting user privacy and it reduces the possibility of privacy breach. 
FL has several additional benefits, such as reduced computation (e.g., preprocessing data, training models) and storage (e.g., supervised or unsupervised data, model versions) needs associated with centralized training, possible model personalization \cite{tan2022towards}, and accelerated convergence \cite{mishchenko2022proxskip}. The prior works in FL mostly draw inferences and propose solutions and best practices based on smaller and/or simpler models (shallow CNNs, Transformers) applied to problems such as classification and regression. There has been a limited emphasis on studying FL for ASR, and thus a limited understanding of the applicability of existing research directly towards ASR that uses much larger Neural Networks (NNs) than most conventional ones studied in the literature. Most existing works in FL for ASR either limit themselves only to preliminary explorations or do not consider various aspects that would be relevant when training ASR models in a practical FL setting, such as data heterogeneity, impact of optimizer, and model~design.

\vspace{1mm}
In this work, we aim to bridge these gaps and provide an in-depth practical guide to training E2E CTC-AED models using FL. The main contributions of this work are as follows:
\begin{itemize}[leftmargin=3.5mm]
    \vspace{-2.5mm}
    \item We examine the applicability of the algorithms, trends, and best practices from prior works, e.g., impact of adaptive optimizers \cite{reddi2021adaptive}, seed start \cite{gao2022end}, etc., in the context of FL for E2E ASR  with CTC-AED loss (\flforasr).
    \vspace{-2.5mm}
    \item We present the choice of model and training parameters that play a pivotal role in bridging the gap between the performance of {\flforasr} models and their centralized counterpart, thus taking a step towards practical {\flforasr} training. These are based on extensive ablation studies to understand the importance of the parameters in isolation. 
    \vspace{-2.5mm}
    \item We also present a fresh perspective (confirmed empirically) that explains why some optimizers work better than others for {\flforasr} training, i.e., by inducing \textit{smoothness or similarity}\footnote{Referring to Lipschitz smoothness and cosine similarity.} among model updates or decreasing heterogeneity.
\end{itemize}


\section{Related Work}
\label{sec:related}

We can categorize prior works mainly into two subsets: (i)~\textit{generic} FL research, and (ii) FL research specific to ASR.

\vspace{1mm}
\noindent\textbf{Generic FL.} Most FL research focuses on studying aspects of FL for smaller models and conventional machine learning (ML) tasks such as classification and regression. For example, \cite{reddi2021adaptive} advocates for the usage of adaptive optimizers operating on \textit{pseudo-gradients}\footnote{The aggregated model updates ($\vec{\Delta}{}{(t)}$ in Alg.~\ref{alg:fl}) in FL can be considered as (pseudo-) gradients for the central/global loss function, i.e., $\mathrm{F}(\bvec{\theta}{}{})$ in~\eqref{eqn:fl_formulation}.} to improve convergence speed, but are not evaluated on large models such as Conformer \cite{gulati2020conformer} for ASR. Other works \cite{azam2022recycling, wang2020tackling} study challenges in FL such as tackling client heterogeneity or improving communication efficiency via gradient compression but are limited in evaluation to conventional settings that do not include large models or more complex tasks such as ASR.

\vspace{1mm}
\noindent\textbf{FL for ASR.} There are only a handful of prior works that focus on FL for ASR models. Specifically, \cite{guliani2021training} train Recurrent Neural Network Transducer (RNN-T) \cite{graves2013speech} models using FL to understand a quality-cost trade-off. Similarly, \cite{guliani2022enabling} study the effect of Federated Dropout on ASR model training but do not consider heterogeneous data distribution, which is one of the main challenges in FL. \cite{gao2022end} comes closest to our work in that it uses End-to-End ASR models, but is different in that 
{(i) they are \textit{unable to train without seed start}, and (ii) train smaller models that consist of CNN-encoder}
and RNN-decoder.
In contrast, our work studies (i) both training with seed start as well as from scratch {using {\lamb} \cite{You2020Large-lamb} to minimize the effect of heterogeneity on large ASR models using layer-wise adaptive learning rate}, (ii) a larger (120 million parameters) model with Conformer encoder
and Bidirectional Transformer decoder,
and (iii) emphasizes on understanding why the use of adaptive optimizers is crucial for ASR models. We also shed light on the differences in takeaways observed between \textit{generic} FL research and {\flforasr}.


\section{Federated Learning For ASR}
\label{sec:framework}

Consider an FL framework with $\mathcal{K}$ clients. During each central \textit{aggregation round}, the central server samples clients $\mathrm{K}\subset \mathcal{K}$ indexed $\mathrm{1,\hdots,\abs{K}}$ where $\abs{\mathrm{K}}$ is termed \textit{cohort size}. Each client $\mathrm{k}$ has a local dataset $\subsup{\mathcal{D}}{\mathrm{k}}{}$ with $\abs{\mathcal{D}_{\mathrm{k}}}$ samples comprising input feature sequence (from speech utterances) and output label sequence (from transcriptions). The objective of the FL framework is to minimize the global loss function $\mathrm{F(\cdot)}$ given by
\eqn{
    \min_{\bvec{\theta}{}{}\in \mathbb{R}^\mathrm{M}} \left\{\mathrm{F}(\bvec{\theta}{}{}) \triangleq \sum_{\mathrm{k=1}}^{\mathrm{\abs{K}}} \frac{\abs{\mathcal{D}_{\mathrm{k}}}}{\subsup{\sum}{k=1}{\abs{K}}\abs{\mathcal{D}_{\mathrm{k}}}} \mathrm{F}_{\mathrm{k}}(\bvec{\theta}{}{})\right\},
    \label{eqn:fl_formulation}
}

\noindent where $\mathrm{F_k(\bvec{\theta}{}{}) = \sum_{d \in \mathcal{D}_k}f(\bvec{\theta}{}{}; d)}/\abs{\mathcal{D}_{\mathrm{k}}}$ is the loss function at client $\mathrm{k}$, such that $\mathrm{f(\bvec{\theta}{}{}; d)}$ denotes the loss (e.g., CTC-AED)  over the data $\mathrm{d}$ given the model parameters $\bvec{\theta}{}{}$. 

An FL framework optimizes the objective \eqref{eqn:fl_formulation} via engaging a sampled set of clients in local model training. The clients train on their own local datasets and periodically upload their model updates to the server, which aggregates the updates and updates the central model either through averaging conventionally \cite{konevcny2016federated}, or through an adaptive optimizer \cite{reddi2021adaptive}. The updated central model is broadcasted to another sampled set of clients and the process is repeated either for a fixed number of central aggregation rounds or until convergence. The FL algorithm and the corresponding terminology used in the rest of this paper are summarized in Algorithm~\ref{alg:fl}.

\begin{algorithm}[t]
\footnotesize
  \caption{Federated Optimization with Client Sampling}
\label{alg:fl}
\begin{algorithmic}[1]
    \State Initialize model $\bvec{\theta}{}{(0)}$ and \colorbox{yellow}{$\mathsf{central\_optimizer}$}
    \For{\colorbox{yellow}{$\mathsf{aggregation\_rounds}$}~$\mathrm{t = 0}$ to $\mathrm{(T-1)}$}
        \State Sample a subset $\mathrm{K}$ from available clients $\mathcal{K}$ (\colorbox{yellow}{$\mathsf{cohort\_size}$}$= \abs{K}$)
        \For{each client $\mathrm{k\in K}$ \textbf{in parallel}}
            \State Initialize local model $\bvec{\theta}{k}{(t, 0, 0)} \leftarrow \bvec{\theta}{}{(t)}$ and \colorbox{yellow}{$\mathsf{local\_optimizer}$}
            \For{\colorbox{yellow}{$\mathsf{local\_epochs}$} $\mathrm{e=0}$ to $\mathrm{E-1}$}
                \For{$\mathsf{local\_batch}$ $\mathrm{b=0}$ to $(\mathrm{B_k}-1)$}
                    \State Compute gradient estimate $\vec{g}{\mathsf{k}}{\mathsf{(t, e, b)}}$
                    \State $\bvec{\theta}{\mathrm{k}}{\mathrm{(t, e, b+1)}} \leftarrow \mathsf{local\_optimizer}\left(\vec{g}{\mathsf{k}}{\mathsf{(t, e, b)}}\right)$
                \EndFor
            \EndFor
            
            \State $\vec{\Delta}{\mathrm{k}}{\mathrm{(t)}} = \bvec{\theta}{\mathrm{k}}{\mathrm{(t, E-1, B_k-1)}} - \bvec{\theta}{\mathrm{k}}{\mathrm{(t, 0, 0)}}$
        \EndFor
        \State $\vec{\Delta}{}{\mathrm{(t)}} = \frac{1}{\abs{K}} \sum_{k\in K} \vec{\Delta}{\mathrm{k}}{\mathrm{(t)}}$
        \State $\bvec{\theta}{}{(0)} \leftarrow \mathsf{central\_optimizer} \left( \vec{\Delta}{}{\mathrm{(t)}} \right)$
    \EndFor
\end{algorithmic}
\end{algorithm}


\vspace{-0.1cm}
\section{Experiments}
\label{sec:expts}

\noindent\textbf{Dataset.} We run FL simulations using LibriSpeech \cite{panayotov2015librispeech}, which consists of over 960 hours of training data divided into 3 training splits and 4 evaluation splits. The statistics about the data splits are detailed in Table~\ref{table:librispeech}. Each data sample comprises audio, its transcript, and the associated speaker ID. 

\begin{table}[t!]
\vspace{-0.2cm}
    \footnotesize
    \centering
    \resizebox{0.8\columnwidth}{!}{
    \begin{tabular}{| c | c | c | c | c | c |}
      \hline
      \multirow{2}{*}{\textbf{split}} & \multirow{2}{*}{\textbf{\# hrs}} &  \multirow{2}{*}{\textbf{\# utts}} & \multirow{2}{*}{\textbf{\# spk}} & \multirow{2}{*}{\textbf{$\frac{\text{min}}{\text{spk}}$}} & \multirow{2}{*}{\textbf{$\frac{\text{utts}}{\text{spk}}$}} \\
      & & & & & \\
      \hline
      \cleanhundred & $100.6$ & $\hphantom{0}28539$ & $\hphantom{0}251$ & $25$ & $\subsub{114}{15}$\\
      \cleanthreesixty & $363.6$ & $104014$ & $\hphantom{0}921$ & $25$ & $\subsub{113}{16}$\\
      \otherfivehundred & $496.7$ & $148688$ & $1166$ & $30$ & $\subsub{128}{30}$ \\\hline
      \devclean & $\hphantom{00}5.4$ & $\hphantom{00}2703$ & $\hphantom{00}40$ & $\hphantom{0}8$& $\subsub{\hphantom{0}68}{17}$\\
      \devother & $\hphantom{00}5.3$ & $\hphantom{00}2864$ & $\hphantom{00}33$ & $10$ & $\subsub{\hphantom{0}87}{25}$ \\
      \testclean & $\hphantom{00}5.4$ & $\hphantom{00}2620$ & $\hphantom{00}40$ & $\hphantom{0}8$ & $\subsub{\hphantom{0}66}{20}$ \\
      \testother & $\hphantom{00}5.1$ & $\hphantom{00}2939$ & $\hphantom{00}33$ & $10$ & $\subsub{\hphantom{0}89}{26}$ \\\hline
    \end{tabular}
    }
    \vspace{-2mm}
    \caption{Statistics about number of hours (\# hrs), utterances (\# utts), unique speakers (\# spk), minutes per speaker ($\frac{\mathrm{min}}{\mathrm{spk}}$), and utterances per speaker ($\frac{\mathrm{utts}}{\mathrm{spk}}$) for LibriSpeech data splits.}
    \label{table:librispeech}
  \vspace{-5mm}
\end{table}

\vspace{0.5mm}
\noindent\textbf{Federated Dataset.} We use the speaker IDs to partition the data so that each speaker corresponds to one client~participating in FL, thus simulating a 
non-independent and identically distributed (non-IID) data distribution among clients in FL. 

\vspace{0.5mm}
\noindent\textbf{Federated Training.} For FL, we consider two different types of training setups: (i) {\flrandomstart} that starts FL training from a randomly initialized model and (ii) {\flseedstart} that starts the training from a centrally trained seed model. For {\flrandomstart}, we train on three different data splits: (i) {\cleanhundred}, (ii) {\cleanhundred~+~\cleanthreesixty~=~\cleanfoursixty}, and (iii) {\cleanfoursixty~+ \otherfivehundred~=~\allninesixty}.  For {\flseedstart}, we split the data into two parts: (i) {\cleanhundred} used for seed model training using centralized training and (ii) {\cleanthreesixty} or {\cleanthreesixty~+~\otherfivehundred~=~\mixedeightsixty} partitioned using speaker IDs used as client data for FL. The {\flseedstart} is realistic since oftentimes there is a small set of data (e.g., open-source or legacy data) available on the server \cite{gao2022end, diao2022semifl}. 

\vspace{0.5mm}
\noindent\textbf{Tokenizer.} We train the SentencePiece \cite{kudo2018sentencepiece} tokenizer with Byte-Pair Encoding (BPE) \cite{sennrich2016neural} subword units only on the {\cleanhundred} split since in a realistic setting we might not have access to all client data on the server to train a tokenizer.

\vspace{0.5mm}
\noindent\textbf{Model.} We use a CTC-AED model similar in size\footnote{Link to the configuration file: \url{https://github.com/wenet-e2e/wenet/blob/main/examples/librispeech/s0/conf/train_conformer_bidecoder_large.yaml}} to the conformer-bidecoder-large WeNet \cite{yao2021wenet} model, which consists of a conformer encoder with 12 blocks each comprising 8 attention heads, and a bidirectional transformer decoder comprising 3 blocks of 8 attention heads in both the left and right decoders. The model consists of $\sim 120$ million parameters.

\vspace{0.5mm}
\noindent\textbf{Training.} We use the the CTC-AED loss for training and use {\attentionrescore} \cite{yao2021wenet}  decoding with $\mathrm{beam\_size}=50$ for evaluation. We train with $8\times 32$~GB V$100$ GPU and batch size of $16$ on each client during FL and $48$ for centralized training, while the test-time inference is done with a larger batch size. For centralized training, we train with data augmentation \cite{park2019specaugment,wu2021u2++}, warmup and learning rate decay scheduler for $80,000$ model updates or until convergence. The training parameters such as {\cohortsize}, {\aggrounds}, {\localepochs}, etc. are reported individually for each experiment.

\vspace{0.5mm}
\noindent\textbf{Evaluation.} The FL experiments report both mean and standard deviation ({\stddev}) of Word Error Rates (as $\wer_{_{\mathrm{std}}}$) on the evaluation sets of LibriSpeech, calculated from at least 3 runs started with different random seeds.

\begin{table}[t!]
    \footnotesize
    \centering
    \begin{tabular}{| c | c | c | c | c | c |}
      \hline
      \multirow{2}{*}{\textbf{baseline}} & \textbf{train} & \multicolumn{2}{c|}{\textbf{dev-}} & \multicolumn{2}{c|}{\textbf{test-}} \\\cline{3-6}
      & \textbf{data}  & \textbf{clean} & \textbf{other} & \textbf{clean} & \textbf{other} \\
      \hline
      {central training} & \cleanhundred & $10.32$ & $26.21$ & $10.18$ & $24.99$ \\
      & \cleanfoursixty  & $\hphantom{0}3.91$ & $12.47$ & $\hphantom{0}3.46$ & $12.54$ \\
      & \allninesixty & $\hphantom{0}{2.79}$ & $\hphantom{0}{7.94}$ & $\hphantom{0}2.96$ & $\hphantom{0}{8.14}$ \\
      \hline\
      {guliani et al.}\cite{guliani2021training}  & \allninesixty & $\hphantom{0}4.60$ & $11.40$ & $\hphantom{0}4.80$ & $11.40$\\
      \hline
    \end{tabular}
    \vspace{-1mm}
    \caption{\textit{Baselines using {\allninesixty} LibriSpeech.} Central training is the achievable performance limit using FL, \cite{guliani2021training} is a prior SOTA that trains an RNN-T on LibriSpeech using FL.}
    \label{table:baselines}
  \vspace{-5mm}
\end{table}

\vspace{0.5mm}
\noindent \textbf{Baselines.} To better understand the impact of various parameters on {\flforasr} training pipeline, we choose two baselines for comparison: (i) a centrally trained model that serves as the limit of achievable performance with FL and (ii) \cite{guliani2021training} which is the state-of-the-art (SOTA) performance on LibriSpeech using FL and non-IID data. We use the $\wer$ as reported in \cite{guliani2021training} owing to the lack of implementation details in the paper.

\begin{table*}[t!]
    \footnotesize
    \newcounter{expt}[table]
    \renewcommand*\theexpt{\thetable.\arabic{expt}}
    \newcommand*\nextexpt[1]{\refstepcounter{expt}\theexpt\label{#1}}
    \centering
    \resizebox{\textwidth}{!}{
    \begin{tabular}{| l | c | c | c | c | c | c | c | c | c | c | l | l | l | l |}
      \hline
      \multirow{2}{*}{\textbf{~id}} & {\textbf{train}} &  \textbf{cohort} & \textbf{local} & \multirow{2}{*}{\textbf{scheduler}} & \textbf{spec} & \textbf{spec} & \multirow{2}{*}{\textbf{c-opt}} & \textbf{seed} & \textbf{opt.} & {\textbf{AED}} & \multicolumn{2}{c|}{\textbf{dev-}} & \multicolumn{2}{c|}{\textbf{test-}} \\\cline{12-15}
      & \textbf{data} & \textbf{size} & \textbf{epochs} & & \textbf{aug} & \textbf{sub} & & \textbf{start} & \textbf{switch} & \textbf{loss} & \multicolumn{1}{c|}{\textbf{clean}} & \multicolumn{1}{c|}{\textbf{other}} & \multicolumn{1}{c|}{\textbf{clean}} & \multicolumn{1}{c|}{\textbf{other}} \\
      \hline
      {\setcounter{table}{1}\setcounter{expt}{0}}\nextexpt{t3r:1-01}  & \multirow{10}{*}{\rotninety{\cleanhundred}} & 5 & 10 & \xmark & \xmark & \xmark & \sgd & \xmark & \xmark & \cmark & $\subsub{30.06}{5.96}$ & $\subsub{52.18}{5.09}$ & $\subsub{30.08}{5.87}$ & $\subsub{53.44}{4.63}$ \\
      \nextexpt{t3r:1-02} & & 5 & 10 & \warmup & \xmark & \xmark & \sgd & \xmark & \xmark & \cmark & $\subsub{28.09}{1.35}$ & $\subsub{51.02}{1.03}$ & $\subsub{28.15}{1.25}$ & $\subsub{52.43}{0.85}$ \\
      \nextexpt{t3r:1-03} & & 5 & 10 & \warmup +\decay & \xmark & \xmark & \sgd & \xmark & \xmark & \cmark & $\subsub{28.20}{0.52}$ & $\subsub{51.53}{0.30}$ & $\subsub{28.14}{0.69}$ & $\subsub{52.92}{0.38}$ \\
      \nextexpt{t3r:1-04} & & 5 & 10 & \warmup +\decay & \cmark & \xmark & \sgd & \xmark & \xmark & \cmark & $\subsub{21.75}{2.93}$ & $\subsub{38.95}{3.50}$ & $\subsub{21.90}{2.87}$ & $\subsub{40.17}{3.43}$ \\
      \nextexpt{t3r:1-05} & & 5 & 10 & \warmup +\decay & \cmark & \cmark & \sgd & \xmark & \xmark & \cmark & $\subsub{20.27}{0.40}$ & $\subsub{36.84}{1.02}$ & $\subsub{20.49}{0.30}$ & $\subsub{38.31}{0.82}$ \\
      \nextexpt{t3r:1-06-1} & & 5 & 10 & \warmup +\decay & \cmark & \cmark & \sgd & \xmark & \xmark & \xmark & $\subsub{62.08}{43.5}$ & $\subsub{70.14}{34.2}$ & $\subsub{62.17}{43.4}$ & $\subsub{70.90}{33.4}$ \\\cdashline{12-15}
      \nextexpt{t3r:1-06} & & 5 & 10 & \warmup +\decay & \cmark & \cmark & \adam & \xmark & \xmark & \cmark; \xmark & \multicolumn{4}{c|}{\multirow{1}{*}{\diverges}} \\\cdashline{12-15}
      \nextexpt{t3r:1-06-2} & & 5 & 10 & \warmup +\decay & \cmark & \cmark & \adam & \xmark & \cmark & \cmark & $\subsub{20.81}{1.64}$ & $\subsub{37.52}{2.47}$ & $\subsub{21.04}{1.60}$ & $\subsub{38.93}{2.34}$ \\
      \nextexpt{t3r:1-07} & & 5 & 10 & \warmup +\decay & \cmark & \cmark & \adam & ~\cmark\warmstart & \xmark & \cmark & $\subsub{10.87}{0.29}$ & $\subsub{25.08}{0.74}$ & $\subsub{10.95}{0.29}$ & $\subsub{26.28}{0.58}$ \\
      \nextexpt{t3r:1-09} & & 5 & 10 & \warmup +\decay & \cmark & \cmark & \sgd & ~\cmark\warmstart & \xmark & \cmark & $\subsub{19.95}{0.83}$ & $\subsub{38.51}{1.35}$ & $\subsub{20.10}{0.74}$ & $\subsub{40.19}{1.21}$ \\
      \hline
      \setcounter{table}{2}\setcounter{expt}{0}\nextexpt{t3r:2-01} & \multirow{12}{*}{\rotninety{\cleanfoursixty}} & 5 & 10 & \xmark & \xmark & \xmark & \sgd & \xmark & \xmark & \cmark & $\subsub{22.31}{3.97}$ & $\subsub{44.05}{3.92}$ & $\subsub{22.57}{3.94}$ & $\subsub{45.43}{4.15}$ \\
      \nextexpt{t3r:2-02} & & 5 & 10 & \warmup & \xmark & \xmark & \sgd & \xmark & \xmark & \cmark & $\subsub{21.37}{2.51}$ & $\subsub{43.16}{2.46}$ & $\subsub{21.40}{2.32}$ & $\subsub{44.70}{2.49}$ \\
      \nextexpt{t3r:2-03} & & 5 & 10 & \warmup +\decay & \xmark & \xmark & \sgd & \xmark & \xmark & \cmark & $\subsub{20.74}{1.07}$ & $\subsub{42.56}{1.29}$ & $\subsub{20.95}{1.13}$ & $\subsub{44.16}{1.62}$ \\
      \nextexpt{t3r:2-04} & & 5 & 10 & \warmup +\decay & \cmark & \xmark & \sgd & \xmark & \xmark & \cmark & $\subsub{16.83}{0.25}$ & $\subsub{33.25}{0.32}$ & $\subsub{17.04}{0.18}$ & $\subsub{34.53}{0.81}$ \\
      \nextexpt{t3r:2-05} & & 5 & 10 & \warmup +\decay & \cmark & \cmark & \sgd & \xmark & \xmark & \cmark & $\subsub{17.21}{0.74}$ & $\subsub{33.77}{1.10}$ & $\subsub{17.59}{0.74}$ & $\subsub{35.06}{0.78}$ \\
      \nextexpt{t3r:2-06-1} & & 5 & 10 & \warmup +\decay & \cmark & \cmark & \sgd & \xmark & \xmark & \xmark & $\subsub{18.14}{1.47}$ & $\subsub{34.00}{1.92}$ & $\subsub{18.27}{1.58}$ & $\subsub{35.29}{1.94}$ \\\cdashline{12-15}
      \nextexpt{t3r:2-06} & & 5 & 10 & \warmup +\decay & \cmark & \cmark & \adam & \xmark & \xmark & \cmark; \xmark & \multicolumn{4}{c|}{{\diverges}} \\\cdashline{12-15}
      \nextexpt{t3r:2-06-2} & & 5 & 10 & \warmup +\decay & \cmark & \cmark & \adam & \xmark & \cmark & \cmark & $\subsub{17.39}{0.29}$ & $\subsub{34.03}{0.44}$ & $\subsub{17.70}{0.20}$ & $\subsub{35.55}{0.51}$ \\
      \nextexpt{t3r:2-07} & & 5 & 10 & \warmup +\decay & \cmark & \cmark & \adam & ~\cmark\warmstart & \xmark & \cmark & $\subsub{\hphantom{0}6.43}{0.21}$ & $\subsub{20.20}{0.98}$ & $\subsub{\hphantom{0}6.67}{0.25}$ & $\subsub{20.78}{0.77}$ \\
      \nextexpt{t3r:2-09} & & 5 & 10 & \warmup +\decay & \cmark & \cmark & \sgd & ~\cmark\warmstart & \xmark & \cmark & $\subsub{14.48}{1.87}$ & $\subsub{32.16}{1.79}$ & $\subsub{14.51}{1.79}$ & $\subsub{33.72}{1.92}$ \\
      \nextexpt{t3r:2-08} & & 5 & 10 & \warmup +\decay & \cmark & \cmark & \adam & \cmark\pretrained & \xmark & \cmark & $\subsub{\hphantom{0}5.72}{0.34}$ & $\subsub{19.62}{1.51}$ & $\subsub{\hphantom{0}5.99}{0.35}$ & $\subsub{19.73}{1.63}$ \\
      \nextexpt{t3r:2-10} & & 5 & 10 & \warmup +\decay & \cmark & \cmark & \sgd & \cmark\pretrained & \xmark & \cmark & $\subsub{\hphantom{0}6.55}{0.12}$ & $\subsub{21.35}{0.27}$ & $\subsub{\hphantom{0}6.91}{0.10}$ & $\subsub{22.17}{0.32}$ \\
      \hline
      \setcounter{table}{3}\setcounter{expt}{0}\nextexpt{t3r:3-01} & \multirow{12}{*}{\rotninety{\allninesixty}} & 5 & 10 & \xmark & \xmark & \xmark & \sgd & \xmark & \xmark & \cmark & $\subsub{57.54}{44.6}$ & $\subsub{65.39}{36.2}$ & $\subsub{57.55}{44.5}$ & $\subsub{65.92}{35.4}$ \\
      \nextexpt{t3r:3-02} & & 5 & 10 & \warmup & \xmark & \xmark & \sgd & \xmark & \xmark & \cmark & $\subsub{18.60}{0.70}$ & $\subsub{34.12}{0.65}$ & $\subsub{18.62}{0.57}$ & $\subsub{35.62}{0.91}$ \\
      \nextexpt{t3r:3-03} & & 5 & 10 & \warmup +\decay & \xmark & \xmark & \sgd & \xmark & \xmark & \cmark & $\subsub{19.07}{0.82}$ & $\subsub{34.33}{0.75}$ & $\subsub{18.85}{0.84}$ & $\subsub{35.70}{0.65}$ \\
      \nextexpt{t3r:3-04} & & 5 & 10 & \warmup +\decay & \cmark & \xmark & \sgd & \xmark & \xmark & \cmark & $\subsub{16.60}{0.24}$ & $\subsub{28.52}{0.43}$ & $\subsub{16.62}{0.23}$ & $\subsub{29.47}{0.41}$ \\
      \nextexpt{t3r:3-05} & & 5 & 10 & \warmup +\decay & \cmark & \cmark & \sgd & \xmark & \xmark & \cmark & $\subsub{17.33}{0.46}$ & $\subsub{29.39}{0.80}$ & $\subsub{17.25}{0.39}$ & $\subsub{30.48}{1.16}$ \\
      \nextexpt{t3r:3-06-1} & & 5 & 10 & \warmup +\decay & \cmark & \cmark & \sgd & \xmark & \xmark & \xmark & $\subsub{16.78}{0.27}$ & $\subsub{28.13}{0.27}$ & $\subsub{16.65}{0.30}$ & $\subsub{29.32}{0.11}$ \\\cdashline{12-15}
      \nextexpt{t3r:3-06} & & 5 & 10 & \warmup +\decay & \cmark & \cmark & \adam & \xmark & \xmark & \cmark; \xmark & \multicolumn{4}{c|}{{\diverges}} \\\cdashline{12-15}
      \nextexpt{t3r:3-06-2} & & 5 & 10 & \warmup +\decay & \cmark & \cmark & \adam & \xmark & \cmark & \cmark & $\subsub{16.65}{0.41}$ & $\subsub{29.24}{0.46}$ & $\subsub{16.60}{0.34}$ & $\subsub{30.30}{0.45}$ \\
      \nextexpt{t3r:3-07} & & 5 & 10 & \warmup +\decay & \cmark & \cmark & \adam & ~\cmark\warmstart & \xmark & \xmark & $\subsub{\hphantom{0}6.24}{0.18}$ & $\subsub{15.38}{0.40}$ & $\subsub{\hphantom{0}6.39}{0.05}$ & $\subsub{16.08}{0.43}$ \\
      \nextexpt{t3r:3-09} & & 5 & 10 & \warmup +\decay & \cmark & \cmark & \sgd & ~\cmark\warmstart & \xmark & \xmark & $\subsub{14.78}{3.27}$ & $\subsub{27.23}{3.86}$ & $\subsub{14.52}{3.28}$ & $\subsub{28.67}{3.50}$ \\
      \nextexpt{t3r:3-08} & & 5 & 10 & \warmup +\decay & \cmark & \cmark & \adam & \cmark\pretrained & \xmark & \xmark & $\subsub{\hphantom{0}5.48}{0.17}$ & $\subsub{15.31}{0.71}$ & $\subsub{\hphantom{0}5.72}{0.16}$ & $\subsub{15.86}{0.69}$ \\
      \nextexpt{t3r:3-10} & & 5 & 10 & \warmup +\decay & \cmark & \cmark & \sgd & \cmark\pretrained & \xmark & \xmark & $\subsub{\hphantom{0}6.65}{0.23}$ & $\subsub{18.97}{0.61}$ & $\subsub{\hphantom{0}6.86}{0.22}$ & $\subsub{19.52}{0.82}$ \\
      \hline
    \end{tabular}
    }
    \setcounter{table}{2}
    \caption{Effect (in terms of $\wer_{_{\mathrm{std}}}$) of learning rate scheduler: warmup (\warmup) and decay schedule (\decay), data augmentation: SpecAugment \cite{park2019specaugment} and SpecSub \cite{wu2021u2++}, central optimizer (c-opt), seed start: warm-start ({\warmstart}), i.e., unconverged (WER = $98.36\%$ on \devclean) and pre-trained ({\pretrained}), i.e., pre-trained (WER = $10.32\%$ on \devclean) FL initialization, \textit{optimizer} (opt.) \textit{switch}, and $\mathrm{AED}$ loss on CTC-AED model training using FL with \aggrounds$=2000$.}
    \label{table:hyperparameter_tune}
  \vspace{-0.4cm}
\end{table*}

\subsection{Mitigating Diverging Model Training}
We start our exploration by examining the convergence of {\flforasr} training. For this, we use Stochastic Gradient Descent ({\sgd}) as the local optimizer and tune over the two central optimizers: (i) {\adam} \cite{kingma2014adam} as suggested in \cite{reddi2021adaptive} and (ii) {\sgd} as suggested in \cite{konevcny2016federated}. In order to obtain a converging model it is important to use learning rate scheduler and data augmentation \cite{park2019specaugment, wu2021u2++} as can be seen in Table~\ref{table:hyperparameter_tune} (rows \ref{t3r:1-01}-\ref{t3r:1-05}, \ref{t3r:2-01}-\ref{t3r:2-05}, \ref{t3r:3-01}-\ref{t3r:3-05}). The main takeaways are as follows: 
\begin{itemize}[leftmargin=3.5mm]
    \vspace{-2mm}
    \item While a \textit{warmup schedule} for local learning rate (LR) improves both $\wer$ and {\stddev} on smaller datasets (see \ref{t3r:1-01}, \ref{t3r:2-01} vs. \ref{t3r:1-02}, \ref{t3r:2-02} respectively), it \textit{significantly} benefits from the larger training data (see \ref{t3r:3-01} vs \ref{t3r:3-02}). LR \textit{decay schedule} further stabilizes training (see {\stddev} in \ref{t3r:1-02}, \ref{t3r:2-02} vs. \ref{t3r:1-03}, \ref{t3r:2-03} respectively). However, prior research \cite{reddi2021adaptive} finds that the LR scheduler only has a ``modest'' effect on performance. 
    \vspace{-2mm}
    \item \textit{Data augmentation} improves $\wer$: SpecAugment \cite{park2019specaugment} leads to \textit{significant} improvements irrespective of training data size (see \ref{t3r:1-03}, \ref{t3r:2-03}, \ref{t3r:3-03} vs. \ref{t3r:1-04}, \ref{t3r:2-04}, \ref{t3r:3-04} respectively) but SpecSub \cite{wu2021u2++} shows visible benefits (including {\stddev}) only for smaller data, i.e., {\cleanhundred} (see \ref{t3r:1-04} vs. \ref{t3r:1-05}).
\end{itemize}
\noindent These observations correspond to a \textit{reduction in heterogeneity among client updates at various stages of training}: warmup and decay scheduler for LR ensure smaller LR at the start and end of training respectively and data augmentation prevents model overfitting to local data, thus reducing heterogeneity among of model updates from client in FL \cite{reddi2021adaptive,yuan2022addressing}. It is worth noting that among \textit{pre-layer norm} and \textit{post-layer norm}, only the former \textit{consistently converges} on all training data sizes using FL. Also, during our initial experiments summarized in Table~\ref{table:hyperparameter_tune}, we observe that {\sgd} outperforms {\adam} when using default parameters, i.e., $\varepsilon=1\mathrm{e}-8$. In fact, {\adam} fails to converge with default parameters as can be seen in Table~\ref{table:hyperparameter_tune}.

\subsection{How to Stabilize Adam Training?}
While tuning epsilon $\varepsilon$ in {\adam} in centralized training affects both the numerical stability and the magnitude of model updates \cite{liu2020On} during optimization, we observe diverging training for all values of $\varepsilon \in \{0.1, 0.01, 1\mathrm{e}-5, 1\mathrm{e}-8\}$ (see rows \ref{t3r:1-06},\,\ref{t3r:2-06},\,\ref{t3r:3-06} in Table~\ref{table:hyperparameter_tune}) when using FL.

\vspace{0.5mm}
\noindent\textbf{Seed Start and $\boldsymbol{\varepsilon}$ together help.} If we tune $\varepsilon$ while starting FL from a mildly trained seed model (WER of seed model is $98.36\%$ on \devclean), we observe a prominent effect of $\varepsilon$ on convergence; $\varepsilon=0.01$ works best (rows \ref{t3r:1-07},\,\ref{t3r:2-07},\,\ref{t3r:3-07} in Table~\ref{table:hyperparameter_tune}). Using $\varepsilon=0.01$ in fact helps {\adam} outperform {\sgd}. While we posit that large $\varepsilon$ reduces the magnitude of model updates, thus minimizing the effect of heterogeneity,~the failure of {\adam} training with random initialization could be resulting from either (i) the inability of attention-layers in the decoder to learn, or (ii) irregular estimation of second-order statistics gathered from noisy pseudo-gradients at the start of FL, thus unable to provide useful bias for updates.

\noindent We rule out the former reason by removing the effect of decoder attention layers in optimization. Specifically, we train {\adam} with CTC loss only, i.e., remove AED loss by setting $\alpha=1$ in CTC-AED loss $\mathrm{L}_{\mathrm{CTC-AED}} = \alpha \mathrm{L}_{\mathrm{CTC}} + (1-\alpha) \mathrm{L}_{\mathrm{AED}}$ \cite{yao2021wenet}. This does not improve the diverging training (see rows \ref{t3r:1-06},\,\ref{t3r:2-06},\,\ref{t3r:3-06} in Table~\ref{table:hyperparameter_tune}). However, we can train {\sgd} with CTC loss (see rows \ref{t3r:1-06-1},\,\ref{t3r:2-06-1},\,\ref{t3r:3-06-1} in Table~\ref{table:hyperparameter_tune}).

\vspace{0.5mm}
\noindent\textbf{Role of AED loss.} The decoder in an encoder-decoder architecture of ASR training using CTC-AED loss can be interpreted as a language model. In Table~\ref{table:hyperparameter_tune}, we observe that while the effect of AED loss, and thus a language model, is significant on a small dataset (rows \ref{t3r:1-05} vs \ref{t3r:1-06-1}), it is only marginal on larger datasets (rows \ref{t3r:2-05}, \ref{t3r:3-05} vs \ref{t3r:2-06-1}, \ref{t3r:3-06-1} respectively), possibly because the tokenizer trained on {\cleanhundred} data is not helpful on the highly heterogeneous data in {\otherfivehundred}. We could possibly benefit from a biased decoder \cite{jain20contextual} on clients, i.e., \textit{weighted aggregate of local and central weights} but biased towards the former. We defer this to future exploration.

\vspace{0.5mm}
\noindent\textbf{Optimizer Switch (\optswitch) and Cohort Warmup.} During {\flrandomstart} training, we {try} (i) ``\textit{optimizer switch}'': start training with {\sgd} as central optimizer for a relatively small number ($\sim 100$) of {\aggrounds} and thereafter switch to {\adam}{, and (ii) ``\textit{cohort warmup}'': start with \cohortsize~$=1$ and slowly ramp-up the \cohortsize} to help stabilize training with {\adam} {by helping estimate more stable second-order statistics. While cohort warmup does not help,} {\optswitch} indeed stabilizes {\flrandomstart} training with {\adam} but only able to emulate {\sgd} performance (rows \ref{t3r:1-05},\,\ref{t3r:2-05},\,\ref{t3r:3-05} vs \ref{t3r:1-06-2},\,\ref{t3r:2-06-2},\,\ref{t3r:3-06-2} respectively in Table~\ref{table:hyperparameter_tune}).

\vspace{0.5mm}
\noindent\textbf{Pre-trained Seed Start.}~We observe benefits (see rows~\ref{t3r:2-08}, \ref{t3r:3-08} in Table~\ref{table:hyperparameter_tune}) when initializing FL with a seed model {(\flseedstart)}~trained centrally on server dataset mutually exclusive from FL~dataset. {{\flseedstart} does not require explicit stabilizers such as {\optswitch} to converge.}

\subsection{Significance of Bias Reduction via Optimizers}
{
One of the main challenges in FL is handling heterogeneous data. In the case of ASR, the heterogeneity can be w.r.t both acoustic features of the speaker (input space), and token frequency in the transcripts (output space). Also, the depth of the model only aggravates this problem. Therefore, observing the diminishing returns from tuning {\adam} optimizer which is known for using biased second-order squared estimates, we turn to other optimizers, namely \adamw, \yogi, and \lamb, that help alleviate some of its associated problems. Table~\ref{table:optimizers} summarizes the key differences among these optimizers. 
\begin{table}[h!]
    \footnotesize
    \centering
  \resizebox{0.92\columnwidth}{!}{
    \begin{tabular}{| c | c | c | c | c | c |}
      \cline{2-6}
      \multicolumn{1}{c|}{} & \multicolumn{5}{c|}{\textbf{optimizers}} \\
      \cline{2-6}
      \multicolumn{1}{c |}{\hphantom{0}} & {\sgd} & {\adam} & {\adamw} & {\yogi} & {\lamb} \\
      \cline{2-6}\hline
      {Adaptive LR} & \xmark & \cmark & \cmark & \cmark & \cmark \\\hline
      {weight decay} & \xmark & \xmark & \cmark & \xmark & \xmark \\\hline 
      {bias correction} & \xmark & \xmark & \xmark & \cmark & \xmark \\\hline
      {layer-based} & \multirow{2}{*}{\xmark} & \multirow{2}{*}{\xmark} & \multirow{2}{*}{\xmark} & \multirow{2}{*}{\xmark} & \multirow{2}{*}{\cmark} \\
      {normalization} & & & & & \\\hline
      {more memory} & \xmark & \cmark & \cmark & \cmark & \cmark \\
      \hline
    \end{tabular}
  }
    \caption{Summary of differences between optimizers.}
    \label{table:optimizers}
  \vspace{-0.2cm}
\end{table}

\noindent For both {\flrandomstart} and {\flseedstart} training, it can be immediately observed that while {\yogi} and {\adamw} improve upon {\adam} as central optimizers, {\lamb} shows significant benefits over {\sgd} as local optimizer (see Table~\ref{table:central_optimizers}). Furthermore, \textit{local optimization via {\lamb} minimizes the performance gap between adaptive central optimizers}.
}
\begin{table}[t!]
    \footnotesize
    \centering
  \resizebox{\columnwidth}{!}{%
    \begin{tabular}{| c | c | c | c | c | c | c | c |}
      \cline{5-8}
      \multicolumn{4}{c|}{} & \multicolumn{4}{c|}{\textbf{cohort\_size~=~5}} \\\cline{2-8}
      \multicolumn{1}{c|}{} & \textbf{train} & \multicolumn{2}{|c|}{\textbf{optimizer}} & \multicolumn{2}{|c|}{\textbf{dev-}} & \multicolumn{2}{|c|}{\textbf{test-}} \\\cline{3-8}
      \multicolumn{1}{c|}{} & \textbf{data} & \textbf{central} & \textbf{local} & \textbf{clean} & \textbf{other} & \textbf{clean} & \textbf{other} \\
      \hline
      \multirow{32}{*}{\rotninety{\textbf{aggregation rounds = 2000}}} & \multicolumn{7}{|c|}{{\flseedstart} with model pre-trained (\cmark\pretrained) on {\cleanhundred} split}\\\cline{2-8}
      & \multirow{6}{*}{\rotninety{\cleanfoursixty}} & \multirow{2}{*}{\adam} & {\sgd} & $\subsub{6.24}{0.14}$ & $\subsub{20.25}{0.79}$ & $\subsub{6.51}{0.10}$ & $\subsub{20.67}{0.92}$ \\
      & & & {\lamb} & \colorbox{yellow}{$\subsub{4.95}{0.08}$} & $\subsub{16.57}{0.34}$ & $\subsub{5.23}{0.09}$ & $\subsub{16.59}{0.40}$ \\\cline{3-8}
      & & \multirow{2}{*}{\adamw} & {\sgd} & $\subsub{5.68}{0.13}$ & $\subsub{18.58}{0.71}$ & $\subsub{6.04}{0.03}$ & $\subsub{18.86}{0.73}$ \\
      & & & {\lamb} & $\subsub{5.34}{0.14}$ & $\subsub{17.73}{0.21}$ & $\subsub{5.75}{0.20}$ & $\subsub{18.00}{0.23}$ \\\cline{3-8}
      & & \multirow{2}{*}{\yogi} & {\sgd} & $\subsub{6.29}{0.13}$ & $\subsub{20.30}{1.00}$ & $\subsub{6.50}{0.16}$ & $\subsub{20.69}{0.94}$ \\
      & & & {\lamb} & $\subsub{4.96}{0.06}$ & $\subsub{16.68}{0.27}$ & $\subsub{5.27}{0.10}$ & $\subsub{16.76}{0.28}$ \\\cline{2-8}
      & \multirow{6}{*}{\rotninety{\allninesixty}} & \multirow{2}{*}{\adam} & {\sgd} & $\subsub{6.42}{0.22}$ & $\subsub{16.73}{0.84}$ & $\subsub{6.44}{0.17}$ & $\subsub{17.01}{0.84}$ \\
      & & & {\lamb} & $\subsub{5.00}{0.13}$ & $\subsub{13.47}{0.27}$ & \colorbox{yellow}{$\subsub{5.18}{0.10}$} & $\subsub{13.44}{0.13}$ \\\cline{3-8}
      & & \multirow{2}{*}{\adamw} & {\sgd} & $\subsub{6.01}{0.26}$ & $\subsub{15.72}{0.54}$ & $\subsub{6.23}{0.17}$ & $\subsub{16.00}{0.70}$ \\
      & & & {\lamb} & $\subsub{5.57}{0.09}$ & $\subsub{14.62}{0.43}$ & $\subsub{5.76}{0.12}$ & $\subsub{15.04}{0.34}$ \\\cline{3-8}
      & & \multirow{2}{*}{\yogi} & {\sgd} & $\subsub{6.36}{0.14}$ & $\subsub{16.57}{0.28}$ & $\subsub{6.53}{0.10}$ & $\subsub{16.91}{0.37}$ \\
      & & & {\lamb} & $\subsub{4.98}{0.02}$ & \colorbox{yellow}{$\subsub{13.50}{0.22}$} & $\subsub{5.21}{0.05}$ & \colorbox{yellow}{$\subsub{13.48}{0.08}$} \\\cline{2-8}
      & \multicolumn{7}{|c|}{{\flrandomstart} \textbf{with} {\optswitch}}\\\cline{2-8}
      & \multirow{6}{*}{\rotninety{\cleanhundred}} & \multirow{2}{*}{\adam} & {\sgd} & $\subsub{20.81}{1.64}$ & $\subsub{37.52}{2.47}$ & $\subsub{21.04}{1.60}$ & $\subsub{38.93}{2.34}$ \\
      & & & {\lamb} & $\subsub{18.19}{0.31}$ & $\subsub{34.54}{0.21}$ & $\subsub{18.36}{0.25}$ & $\subsub{36.18}{0.51}$ \\\cline{3-8}
      & & \multirow{2}{*}{\adamw} & {\sgd} & $\subsub{19.28}{0.25}$ & $\subsub{35.82}{0.12}$ & $\subsub{19.36}{0.22}$ & $\subsub{37.33}{0.11}$ \\
      & & & {\lamb} & $\subsub{18.19}{0.32}$ & $\subsub{34.48}{0.36}$ & $\subsub{18.27}{0.20}$ & $\subsub{36.30}{0.49}$ \\\cline{3-8}
      & & \multirow{2}{*}{\yogi} & {\sgd} & $\subsub{19.13}{0.11}$ & $\subsub{35.74}{0.14}$ & $\subsub{19.34}{0.19}$ & $\subsub{37.24}{0.29}$ \\
      & & & {\lamb} & $\subsub{18.20}{0.39}$ & $\subsub{34.29}{0.41}$ & $\subsub{18.29}{0.44}$ & $\subsub{36.12}{0.58}$ \\\cline{2-8}
      & \multirow{6}{*}{\rotninety{\cleanfoursixty}} & \multirow{2}{*}{\adam} & {\sgd} & $\subsub{17.39}{0.29}$ & $\subsub{34.03}{0.44}$ & $\subsub{17.70}{0.20}$ & $\subsub{35.55}{0.51}$ \\
      & & & {\lamb} & $\subsub{14.14}{0.19}$ & $\subsub{30.82}{0.77}$ & $\subsub{14.28}{0.21}$ & $\subsub{32.59}{0.91}$ \\\cline{3-8}
      & & \multirow{2}{*}{\adamw} & {\sgd} & $\subsub{15.67}{0.22}$ & $\subsub{32.51}{0.30}$ & $\subsub{15.83}{0.17}$ & $\subsub{33.95}{0.47}$ \\
      & & & {\lamb} & $\subsub{14.17}{0.38}$ & $\subsub{30.97}{0.78}$ & $\subsub{14.27}{0.32}$ & $\subsub{32.44}{0.94}$ \\\cline{3-8}
      & & \multirow{2}{*}{\yogi} & {\sgd} & $\subsub{15.62}{0.23}$ & $\subsub{32.36}{0.25}$ & $\subsub{15.79}{0.08}$ & $\subsub{33.73}{0.35}$ \\
      & & & {\lamb} & $\subsub{14.11}{0.15}$ & $\subsub{30.67}{0.13}$ & $\subsub{14.31}{0.10}$ & $\subsub{32.04}{0.26}$ \\\cline{2-8}
      & \multirow{6}{*}{\rotninety{\allninesixty}} & \multirow{2}{*}{\adam} & {\sgd} & $\subsub{16.65}{0.41}$ & $\subsub{29.24}{0.46}$ & $\subsub{16.60}{0.34}$ & $\subsub{30.30}{0.45}$ \\
      & & & {\lamb} & \colorbox{yellow}{$\subsub{13.73}{0.17}$} & $\subsub{26.25}{0.58}$ & \colorbox{yellow}{$\subsub{13.66}{0.23}$} & $\subsub{27.50}{0.68}$ \\\cline{3-8}
      & & \multirow{2}{*}{\adamw} & {\sgd} & $\subsub{15.26}{0.34}$ & $\subsub{27.54}{0.30}$ & $\subsub{15.17}{0.18}$ & $\subsub{28.67}{0.44}$ \\
      & & & {\lamb} & $\subsub{13.81}{0.10}$ & $\subsub{26.01}{0.72}$ & $\subsub{13.72}{0.02}$ & \colorbox{yellow}{$\subsub{27.28}{0.61}$} \\\cline{3-8}
      & & \multirow{2}{*}{\yogi} & {\sgd} & $\subsub{15.33}{0.37}$ & $\subsub{27.73}{0.27}$ & $\subsub{15.40}{0.26}$ & $\subsub{28.94}{0.43}$ \\
      & & & {\lamb} & $\subsub{13.82}{0.15}$ & \colorbox{yellow}{$\subsub{26.01}{0.60}$} & $\subsub{13.70}{0.15}$ & $\subsub{27.38}{0.50}$ \\\hline
    \end{tabular}
  }
    \caption{{\adam}, {\adamw}, and {\yogi} as central optimizer and {\sgd} and {\lamb} as local optimizer for {\flseedstart} and {\flrandomstart} with {\optswitch} (\localepochs$=10$). }
    \label{table:central_optimizers}
  \vspace{-0.2cm}
\end{table}

\begin{table}[t!]
    \footnotesize
    \centering
  \resizebox{\columnwidth}{!}{%
  {
    \begin{tabular}{| c | c | c | c | c | c | c |}
      \cline{4-7}
      \multicolumn{3}{c|}{}  & \multicolumn{2}{c|}{\textbf{cohort\_size~=~10}} & \multicolumn{2}{c|}{\textbf{cohort\_size~=~15}} \\\cline{2-7}
      \multicolumn{1}{c|}{} & \textbf{train} & {\textbf{central}} & \multicolumn{2}{c|}{\textbf{test-}}  & \multicolumn{2}{c|}{\textbf{test-}} \\\cline{4-7}
      \multicolumn{1}{c|}{} & \textbf{data} & \textbf{optimizer} & \textbf{clean} & \textbf{other} & \textbf{clean} & \textbf{other} \\
      \hline
      \multirow{7}{*}{\rotninety{\textbf{agg. rounds = 2000}~}} & \multicolumn{6}{|c|}{{\flseedstart} with model pre-trained (\cmark\pretrained) on {\cleanhundred} split}\\\cline{2-7}
      & \multirow{3}{*}{\rotthirty{\cleanfoursixty}} & {\adam} & $\subsub{4.79}{0.04}$ & $\subsub{15.37}{0.62}$ & $\subsub{4.69}{0.04}$ & $\subsub{15.16}{0.19}$\\
      & & {\adamw} & $\subsub{5.11}{0.04}$ & $\subsub{16.29}{0.43}$ & $\subsub{4.83}{0.10}$ & $\subsub{15.20}{0.40}$ \\
      & & {\yogi} & $\subsub{4.77}{0.01}$ & $\subsub{15.59}{0.61}$ & $\subsub{4.66}{0.04}$ & $\subsub{14.96}{0.13}$\\\cline{2-7}
      & \multirow{3}{*}{\rotthirty{\allninesixty}} & {\adam} & $\subsub{4.45}{0.09}$ & $\subsub{12.17}{0.17}$ & $\subsub{4.30}{0.06}$ & $\subsub{11.73}{0.22}$\\
      & & {\adamw} & $\subsub{4.64}{0.06}$ & $\subsub{12.96}{0.17}$ & \colorbox{yellow}{$\subsub{4.25}{0.05}$} & \colorbox{yellow}{$\subsub{11.33}{0.07}$} \\
      & & {\yogi} & $\subsub{4.50}{0.10}$ & $\subsub{12.16}{0.09}$ & $\subsub{4.29}{0.02}$ & $\subsub{11.78}{0.38}$\\\hline
      \multirow{10}{*}{\rotninety{\textbf{agg. rounds = 8000}~}}  & \multicolumn{6}{|c|}{{\flrandomstart} \textbf{without} {\optswitch}}\\\cline{2-7}
      & \multirow{3}{*}{\rotthirty{\cleanhundred}} & {\adam} & $\subsub{12.58}{0.34}$ & $\subsub{26.80}{0.18}$ & $\subsub{13.62}{0.70}$ & $\subsub{27.56}{0.64}$\\
      & & {\adamw} & $\subsub{11.78}{0.23}$ & $\subsub{27.36}{0.26}$ & $\subsub{12.42}{0.12}$ & $\subsub{27.54}{0.27}$ \\
      & & {\yogi} & $\subsub{12.81}{0.45}$ & $\subsub{26.92}{0.52}$ & $\subsub{14.27}{0.21}$ & $\subsub{28.36}{0.33}$\\\cline{2-7}
      & \multirow{3}{*}{\rotthirty{\cleanfoursixty}} & {\adam} & $\subsub{6.58}{0.21}$ & $\subsub{15.45}{0.50}$ & $\subsub{5.95}{0.14}$ & $\subsub{14.32}{0.12}$\\
      & & {\adamw} & $\subsub{5.11}{0.19}$ & $\subsub{15.96}{0.43}$ & $\subsub{4.38}{0.06}$ & $\subsub{14.05}{0.15}$ \\
      & & {\yogi} & $\subsub{6.54}{0.19}$ & $\subsub{15.63}{0.15}$ & $\subsub{6.05}{0.17}$ & $\subsub{14.43}{0.30}$\\\cline{2-7}
      & \multirow{3}{*}{\rotthirty{\allninesixty}} & {\adam} & $\subsub{6.89}{0.27}$ & $\subsub{12.81}{0.31}$ & $\subsub{6.19}{0.40}$ & $\subsub{11.43}{0.35}$\\
      & & {\adamw} & $\subsub{4.98}{0.06}$ & $\subsub{12.69}{0.15}$ & \colorbox{yellow}{$\subsub{3.96}{0.08}$} & \colorbox{yellow}{$\subsub{10.29}{0.22}$} \\
      & & {\yogi} & $\subsub{6.10}{0.10}$ & $\subsub{11.83}{0.27}$ & $\subsub{6.16}{0.46}$ & $\subsub{11.43}{0.23}$\\\hline
    \end{tabular}
  }}
    \caption{{{\adam}, {\adamw}, and {\yogi} as central optimizers and {\lamb} as local optimizer for {\flseedstart} and {\flrandomstart} without {\optswitch} (\localepochs$= 10$). {\optswitch} becomes redundant as we increase {\cohortsize} and comparable performance using {\flrandomstart} can be achieved by increasing the number of \aggrounds. }}
    \label{table:effect_of_cohort_size}
  \vspace{-0.3cm}
\end{table}

\vspace{0.5mm}
\noindent\textbf{Memory Consideration for Local Optimizers.} The choice of local optimizers has an effect on training on edge devices with memory constraints, e.g., compared to vanilla {\sgd}, {\sgd} with momentum requires {2$\times$} and {\lamb} requires {4$\times$} the memory. The exploration of achieving the benefits of adaptive local optimizers while limiting memory consumption would be an interesting future extension.

\subsection{Effect of  Cohort Size and Number of Local Epochs}

\vspace{0.5mm}
\noindent\textbf{Cohort Size.} We examine the effect of increasing the cohort size on ASR performance.
{This leads to an increase in Lipschitz constant of global loss $\var{F}{}{}(\Vec{\theta}{}{})$ in \eqref{eqn:fl_formulation}, thus effect of {\adamw} via inducing smoothness is more prominent~as~seen in Table~\ref{table:effect_of_cohort_size}, especially for {\flrandomstart}. Furthermore, while {\adamw} and {\yogi} (in Table~\ref{table:central_optimizers}; {\cohortsize}~$=5$) can achieve lower {$\wer$} compared to \adam, they have a tendency to sometimes diverge during training without {\optswitch}, which however becomes redundant for larger~cohort sizes in Table~\ref{table:effect_of_cohort_size} since \textit{larger cohort sizes result in second-order estimates that are generalizable across aggregation rounds}, thus improving adaptive scaling of learning rates.
}

\begin{figure}[t!]
    \centering
    \vspace{-0.1cm}
    \includegraphics[width=1.0\columnwidth]{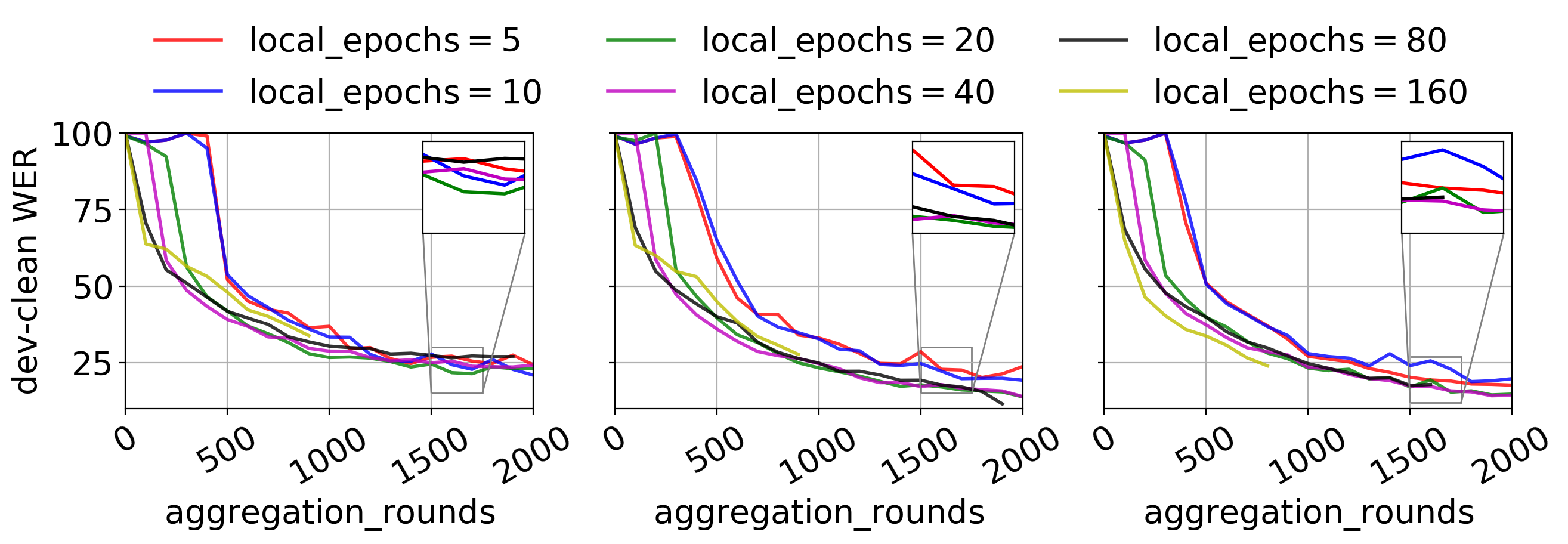}
    \vspace{-0.65cm}
    \caption{The effect of the number of local epochs on the convergence rate and performance for (left) {\cleanhundred}, (middle) \cleanfoursixty, and (right) {\allninesixty} training data. A higher number of local epochs corresponds to faster convergence but gives diminishing returns (see green vs black lines) with a possible plateau on higher WER. This is consistent with observation in \cite{mishchenko2022proxskip}, i.e., diminishing benefits from increasing local epochs.}
    \label{fig:num_of_local_epochs}
    \vspace{-0.3cm}
\end{figure}

\vspace{0.5mm}
\noindent\textbf{Number of Local Epochs.} In Fig.~\ref{fig:num_of_local_epochs}, we observe that changing the number of {\localepochs} in Alg.~\ref{alg:fl} affects both the convergence speed and the final model performance, i.e.,
increasing the number of local epochs (also increases the training time thus early stopping for \localepochs~$=160$) sees a sharper dip in $\wer$ for the same number of aggregation rounds but plateaus quickly at a larger value. A similar observation has been noted in prior work \cite{mishchenko2022proxskip}. This can be used to speed up the training convergence by {dynamically controlling the number of local epochs on clients during training}.


\vspace{-0.4cm}
\section{Discussions}
\label{sec:discuss}

We next study why some optimizers outperform others during FL training by analyzing the overlap among (i) client model updates, (ii) pseudo-gradients, and (iii) central model updates.

\vspace{0.5mm}
\noindent\textbf{Weight Decay in AdamW.} {\adamw} \cite{loshchilov2018decoupled-adamw} introduces an additional weight decay term to the conventional {\adam} optimizer, which we posit helps to increase the smoothness of the underlying gradient subspace. \cite{ning2023input} observed a similar effect of regularization in optimization objective.
Also, since \textit{Lipschitz constant increases exponentially with an increase in the depth of a neural network} \cite{zou2019lipschitz} 
we observe the differences in trends between deep models (such as E2E ASR models) versus the ones conventionally used in FL research \cite{reddi2021adaptive,wang2020tackling}. 

\vspace{0.5mm}
\noindent\textbf{Bias Correction in Yogi.} The main difference between~{\adam} and {\yogi} is the bias correction term introduced in the latter. To understand the effect of the bias correction term,~we~study the overlap (in terms of cosine similarity) among~{central}~model updates at different aggregation rounds. A similar analysis has been used in \cite{azam2022recycling} to develop a compression algorithm that exploits the smoothness of the pseudo-gradient subspace. We observe a similar effect when using {\yogi} instead of {\adam} as seen in Fig.~\ref{fig:model_update_overlaps}. Specifically, the additional term introduced in {\yogi} effectively increases the overlap among model updates for all layers; observed via the diagonal white beam in the second row that is wider for {\yogi} than {\adam}. The smoothing of the pseudo-gradient subspace thus enables the optimizer to minimize the effect of heterogeneity among client updates.
\begin{table}[h]
    \centering
  \resizebox{0.88\columnwidth}{!}{
    \begin{tabular}{| c | c | c |}
      \cline{2-3}
      \multicolumn{1}{c|}{} & \multicolumn{2}{c|}{\textbf{Central Optimizers}} \\\cline{2-3}
      \multicolumn{1}{c|}{} & \textbf{Yogi} & \textbf{Adam} \\
      \cline{1-3}
      \rotninety{\parbox[c]{3.5cm}{\centering \textbf{pseudo-gradient}\\\textbf{similarity grid}}} & 
        \raisebox{-0.5\totalheight}{\includegraphics[width=0.4\linewidth]{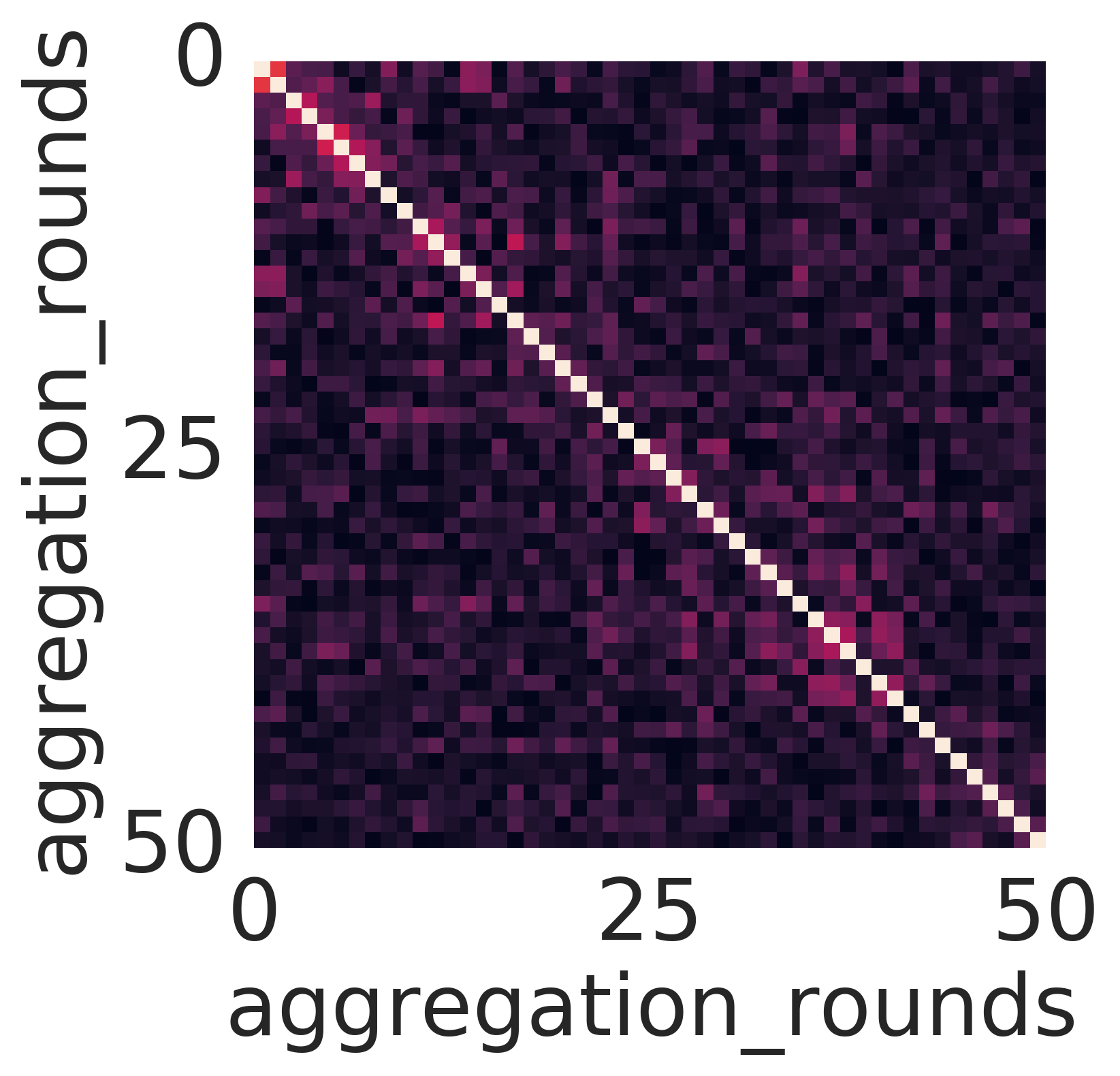}} & 
        \raisebox{-0.5\totalheight}{\includegraphics[width=0.4\linewidth]{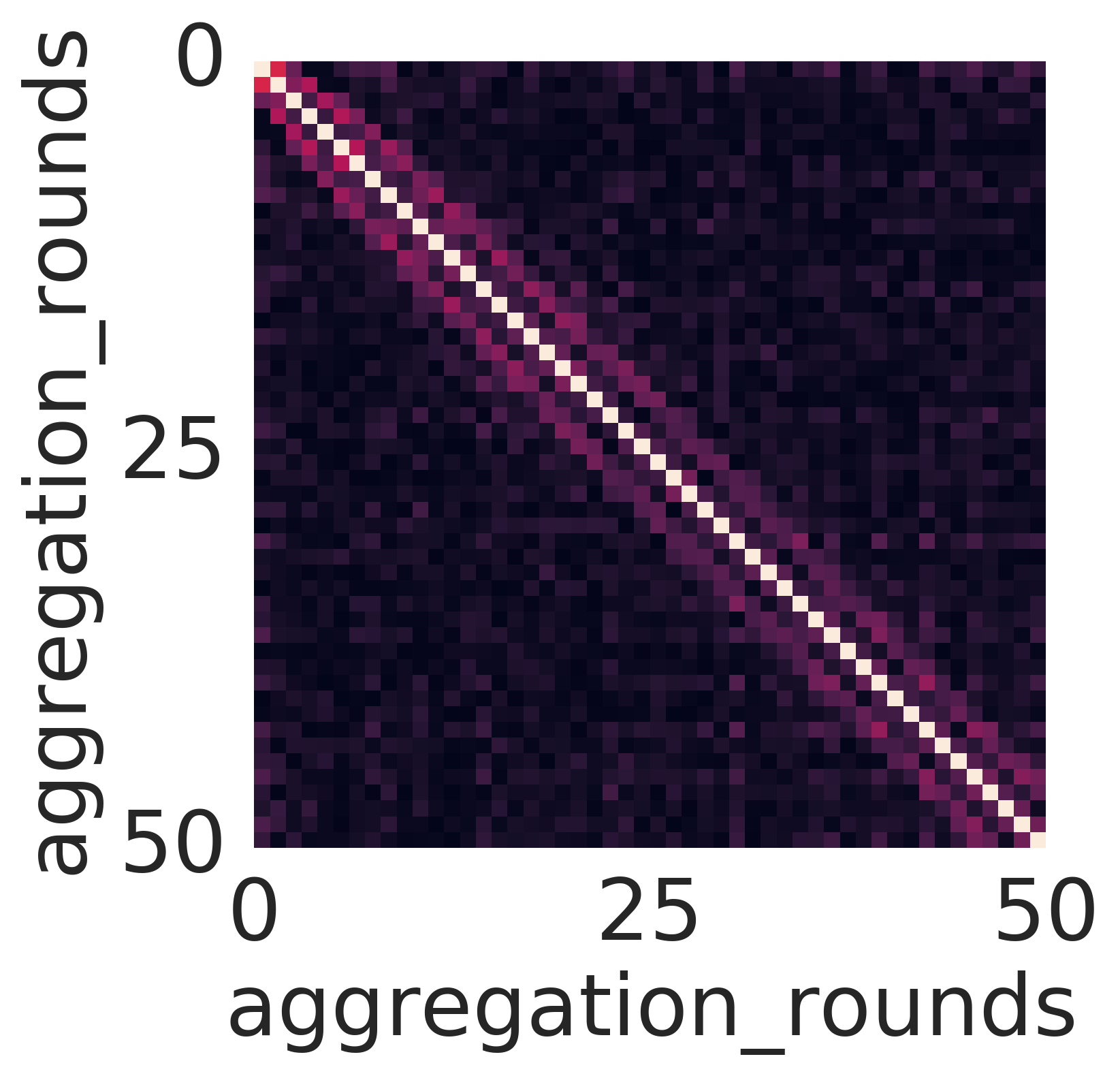}} \\\hline
      \rotninety{\parbox[c]{3.5cm}{\centering \textbf{model update}\\\textbf{similarity grid}}} & 
        \raisebox{-0.5\totalheight}{\includegraphics[width=0.4\linewidth]{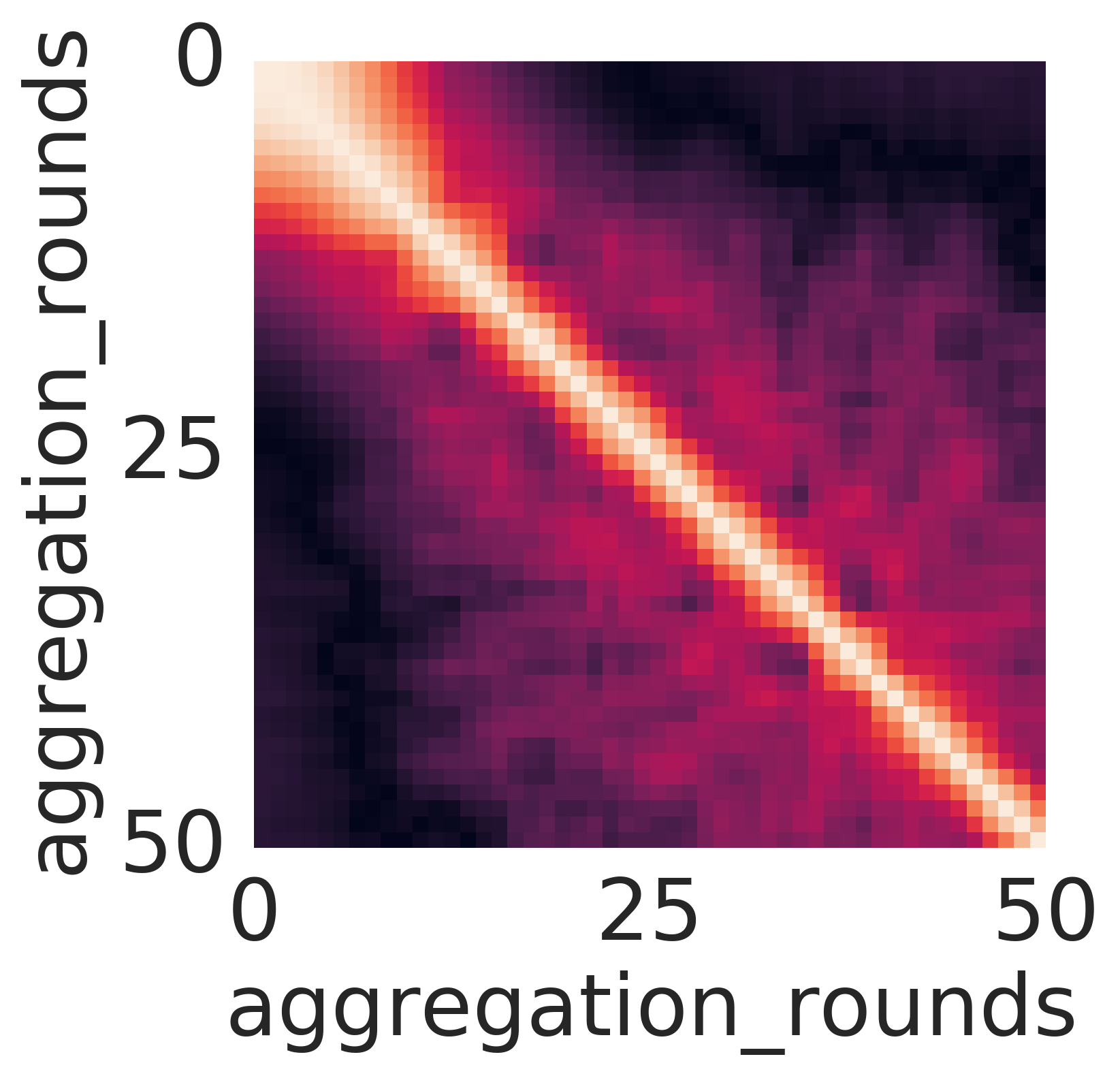}} & 
        \raisebox{-0.5\totalheight}{\includegraphics[width=0.4\linewidth]{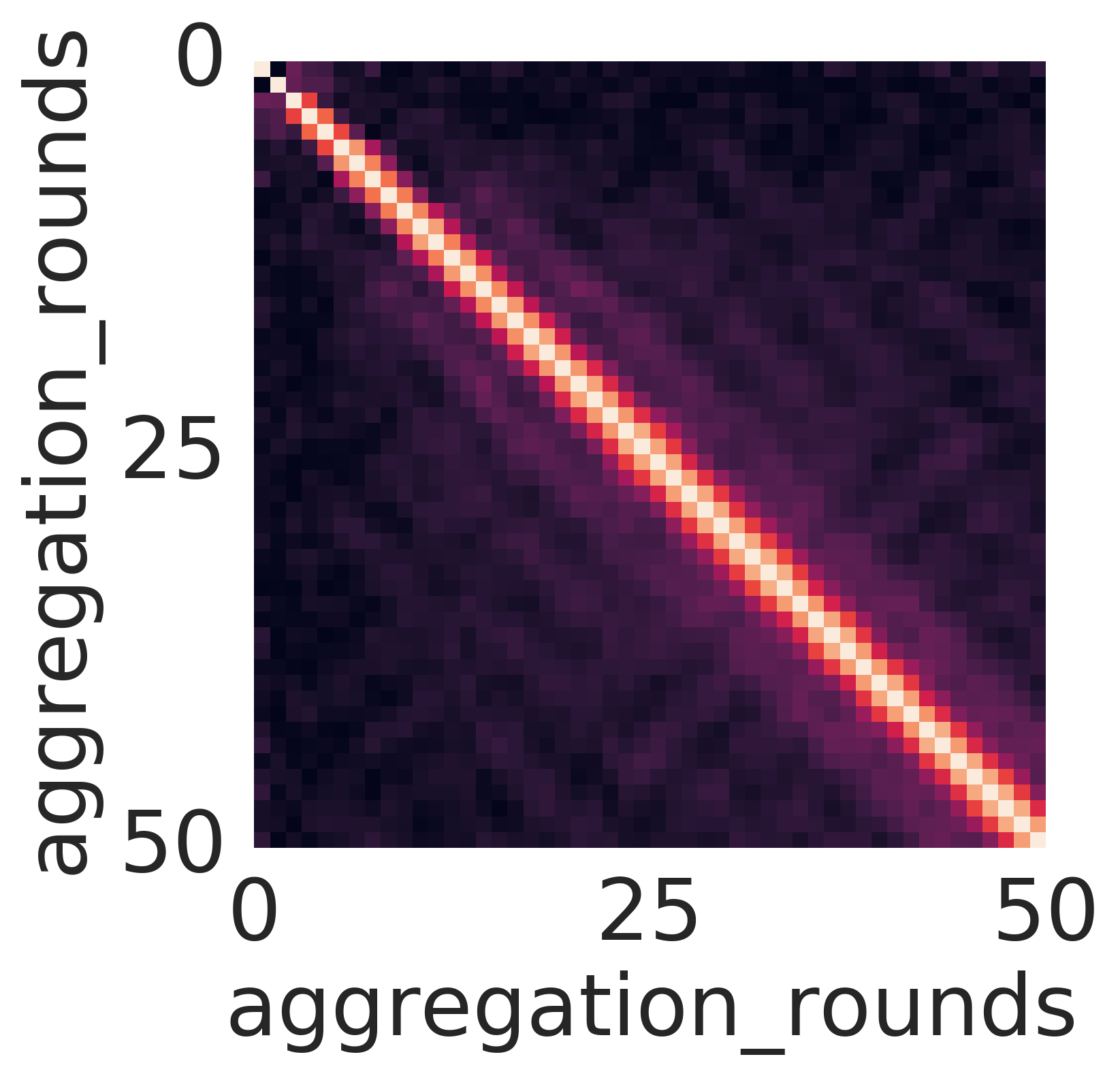}}\\\hline
    \end{tabular}
  }
  \captionof{figure}{Overlap (lighter color means higher cosine similarity) among \textit{central} model updates for {\yogi} \cite{zaheer2018adaptive-yogi} and {\adam} \cite{kingma2014adam} generated for first $50$ aggregation rounds for the $7^{\mathrm{th}}$ encoder layer (conv) with $\sim 3$M parameters in the CTC-AED model.}
  \label{fig:model_update_overlaps}
  \vspace{-0.2cm}
\end{table}

\begin{table}[h]
    \centering
  \resizebox{0.88\columnwidth}{!}{
    \begin{tabular}{| c | c | c |}
      \cline{2-3}
      \multicolumn{1}{c|}{} & \multicolumn{2}{c|}{\textbf{Local Optimizers}} \\\cline{2-3}
      \multicolumn{1}{c|}{} & \textbf{SGD} & \textbf{LAMB} \\
      \hline
      \rotninety{\parbox[c]{3.5cm}{\centering \textbf{client update}\\ \textbf{similarity grid}}} & 
        \raisebox{-0.5\totalheight}{\includegraphics[width=0.4\linewidth]{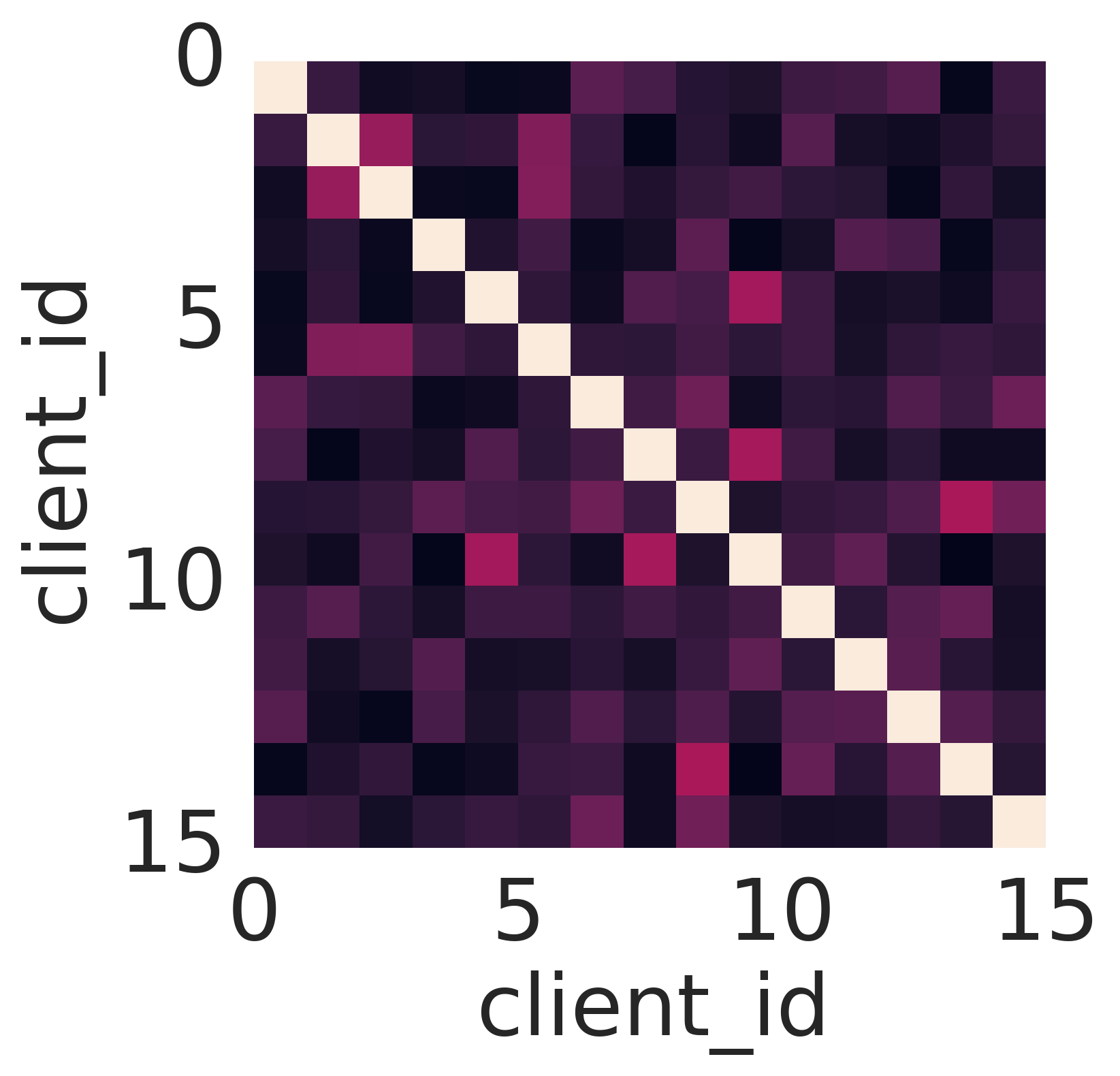}} & 
        \raisebox{-0.5\totalheight}{\includegraphics[width=0.4\linewidth]{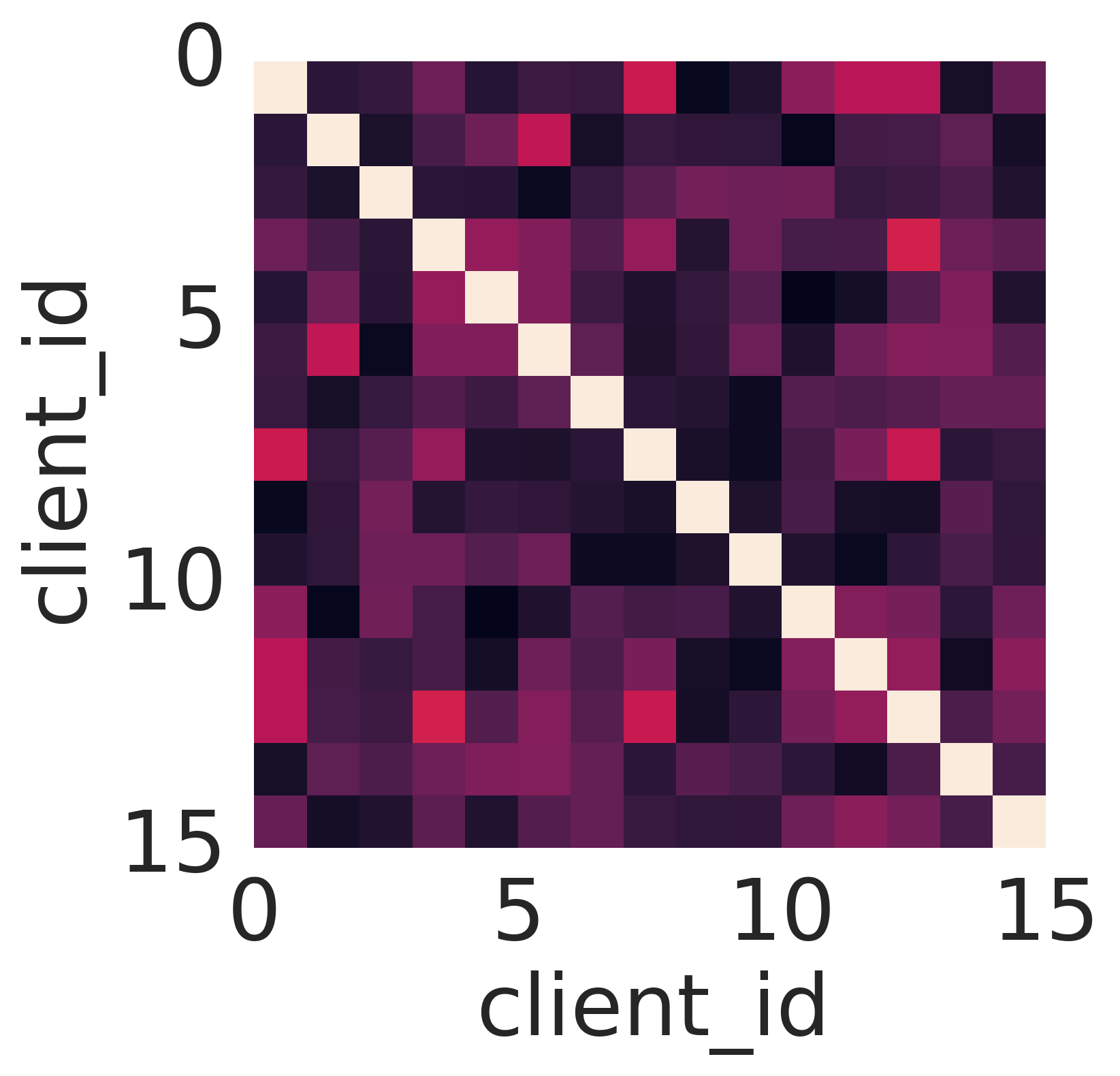}} \\\hline
       \textbf{mean} & 
        $0.14$ & 
        \colorbox{yellow}{$0.23$} \\\hline
    \end{tabular}
  }
  \captionof{figure}{Overlap (lighter color means higher cosine similarity) among \textit{client} model updates for {\sgd} and {\lamb} \cite{You2020Large-lamb} (central optimizer is \yogi) generated for $15$ clients in an aggregation round for the $7^{\mathrm{th}}$ encoder layer (conv) with $\sim 3$ million parameters 
  in the CTC-AED model.}
  \label{fig:client_diff_overlaps}
  \vspace{-0.2cm}
\end{table}

\vspace{0.5mm}
\noindent\textbf{Layer Normalization in LAMB.} The Layer-wise Adaptive Moments Based (\lamb) optimizer \cite{You2020Large-lamb} adds a norm-based adaptive term to {\adam} that mitigates gradient explosion during training with large mini-batches. This property is especially helpful for local training in FL since it effectively lowers the learning rate for larger layers known to generalize poorly because of overfitting \cite{caruana2000overfitting} thus decreasing the heterogeneity among clients as seen via the similarity grid in Fig.~\ref{fig:client_diff_overlaps}. We also observe that the {\lamb} optimizer has a more significant effect when training from a random initialization instead of seed start, thus suggesting that {\flforasr} is more vulnerable to data heterogeneity in the initial phase of training.

\vspace{0.5mm}
\noindent\textbf{Other loss/gradient smoothing methods.} 
We could derive similar benefits from other smoothing techniques, such as label smoothing \cite{muller2019does}, mix-up \cite{zhang2018mixup}, variational dropout \cite{molchanov2017variational}, addition of noise to client model updates \cite{gulati2020conformer,graves2013speech}, however, we leave these explorations to future extensions of this work.


\section{Conclusion}
\label{sec:conclusion}

In this work, we explored various aspects of E2E ASR model training using FL. We summarized various considerations that are essential in minimizing the performance gap between centralized learning and FL. We presented a deep dive into the various aspects of adaptive optimizers that affect optimization during ASR model training, specifically, the impact of bias-correction in the central optimizer, {
reduction of heterogeneity via layer-wise adaptive learning rates}, and thus their significance in FL training. We also explored the effect of seed start on model training, especially its impact on stabilizing training for {\adam} optimizer, as well as other hyperparameters such as number of local epochs and cohort size. Finally, we presented a fresh perspective to explain why some optimizers perform better than others and identified open research directions that would be essential in developing practical FL algorithms.


\clearpage
\section{ACKNOWLEDGMENTS}
\label{sec:ack}

We would like to thank (in alphabetical order) (i) Dan Busbridge, Roger Hsiao, Navdeep Jaitly, and Barry Theobald for the helpful feedback on the initial paper version; (ii) Vitaly Feldman and Kunal Talwar for the helpful discussions on federated learning and experiments throughout the work.

\bibliographystyle{IEEEbib}
\bibliography{references}

\end{document}